\documentclass[11pt,conference,a4paper]{IEEEtran}
\pdfoutput=1
\usepackage{booktabs} 
\usepackage[pdftex]{graphicx}
\usepackage{amsmath}
\usepackage{enumitem}
\usepackage{setspace}
\usepackage{subfig}
\usepackage{listings}
\usepackage{xcolor}
\usepackage{textcomp}

\lstdefinestyle{myc}{
	belowcaptionskip=1\baselineskip,
	breaklines=true,
	xleftmargin=\parindent,
	language=C,
	showstringspaces=false,
	basicstyle=\tiny\ttfamily,
	keywordstyle=\bfseries\color{green!40!black},
	commentstyle=\itshape\color{purple!40!black},
	identifierstyle=\color{blue},
	stringstyle=\color{orange},
}

\lstdefinestyle{myvhdl}{
	belowcaptionskip=1\baselineskip,
	breaklines=true,
	xleftmargin=\parindent,
	language=VHDL,
	showstringspaces=false,
	basicstyle=\tiny\ttfamily,
	keywordstyle=\bfseries\color{green!40!black},
	commentstyle=\itshape\color{purple!40!black},
	identifierstyle=\color{blue},
	stringstyle=\color{orange},
}

\usepackage{cite}
\graphicspath{{./figures/}}

\title{Preliminary Performance Estimations and Benchmark Results for a Software-based Fault-Tolerance Approach aboard Miniaturized Satellite Computers}
\author{Christian M. Fuchs, Todor Stefanov, Nadia Murillo, and Aske Plaat}

\begin{document}
	
	\maketitle
	
	\begin{abstract}
		Modern embedded technology is a driving factor in satellite miniaturization, contributing to a massive boom in satellite launches and a rapidly evolving new space industry.
		Miniaturized satellites however suffer from low reliability, as traditional hardware-based fault-tolerance (FT) concepts are ineffective for on-board computers (OBCs) utilizing modern systems-on-a-chip (SoC).
		Larger satellites therefore continue to rely on proven processors with large feature sizes.
		Software-based concepts have largely been ignored by the space industry as they were researched only in theory, and have not yet reached the level of maturity necessary for implementation.
		In related work, we presented the first integral, real-world solution to enable fault-tolerant general-purpose computing with modern multiprocessor-SoCs (MPSoCs) for spaceflight, thereby enabling their use in future high-priority space missions.
		The presented multi-stage approach consists of three FT stages, combining coarse-grained thread-level distributed self-validation, FPGA reconfiguration, and mixed criticality to assure long-term FT and excellent scalability for both resource constrained and critical high-priority space missions.
		As part of the ongoing implementation effort towards a hardware prototype, several software implementations were achieved and tested.
		This document contains an outline of the conducted tests, performance evaluation results, and supplementary information not included in the actual paper.
		It is being continuously expanded and updated.
	\end{abstract}
	

	\section{Worst-Case Performance Estimation}
	
	To achieve worst-case performance estimations, we developed a naive and unoptimized implementation of the first stage of our approach in C.
	The provided benchmark results were generated based on code derived off a special CCD readout program used for space-based astronomical instrumentation.
	The application was executed with a varying amount of data processing runs in a tile group at the indicated checking frequencies, and without protection for reference.
	
	\subsection{Implementation Outline}
	This implementation was written in approximately 800 lines of user-space C-code including benchmark facilities.
	It utilizes system calls and the POSIX threading library to simulate tiles and thread management.
	Thread-management at this level is computationally expensive, and is drastically faster in a kernel-level MPSoC based implementation.
	Besides enabling very pessimistic benchmarking, this implementation also serves as an excellent simulator to validate the correctness of the described logic, and allows better debugging than on the actual MPSoC implementation.
	Also, this allows to assess the performance of our approach pessimistically without requiring a full prototype in hardware, as simulating such a large FPGA design is only viable for debugging purposes but not for benchmarking.
	
	\subsection{Test Application}

	We encountered severe difficulties trying to benchmark this approach, as synthetic, widely used benchmark suites are unsuitable to benchmark OS-level functionality.
	Thus, we developed we derived a demo-application, off a representative, open-source and well documented scientific computing application.
	We chose to utilize the background scenario of scientific computing, as devices for scientific instrumentation are usually better documented.
	They often do not contain proprietary binary-only black-box applications and sometimes are published as open source \footnote{especially NASA and other governmentally funded projects are required to publish specifications and software open source or in the public domain}.
	Once available, we will also perform benchmarking with a representative command \& data handling application provided by the project sponsor.
	
	The program flow of our demo application is thus based on the NASA/James Webb Space Telescope's Mid-Infrared Instrument (MIRI) described in \cite{ressler2015mid}.
	This program continuously reads three 16-bit 1024x1024 false-color sensor arrays and stores the results.
	It then averages multiple captured frames to optimize the instruments exposure time and avoid saturated pixels or capture faint astronomical sources \cite{ressler2015mid}.
	
	\subsection{Methodology and Test Setup}
	
	The setup simulates three tiles executing the described demo application.
	For each plot in Figure \ref{fig:performance}, 100 measurements were taken of the real-time necessary to process 600 1-Megapixel frames with subsequent processing runs.
	Data heavy modes indicate a high amount of post-processing runs, whereas compute-heavy modes indicate lower per-thread workload.
	
	\begin{itemize}
		\item Very Compute Heavy: 60000 Postprocessing Runs
		\item Compute Heavy: 75000 Postprocessing Runs
		\item Balanced Compute Heavy: 90000 Postprocessing Runs
		\item Balanced Data Heavy: 105000 Postprocessing Runs
		\item Data Heavy: 135000 Postprocessing Runs
		\item Very Data Heavy: 150000 Postprocessing Runs
	\end{itemize}

	Benchmark results were generated on a Intel Core I7 Sandy Bridge-based system with a host kernel's scheduling frequency of 1kHz (\texttt{CONFIG\_HZ\_1000}).
	Hyper-Threading was disable to avoid interference between threads.
	Binaries compiled with \texttt{GCC 6.3.1 (20161221)} without compiler optimization (\texttt{-O0}).
	
	\subsection{Results}

\begin{figure}[!htb]
		\vspace{-15pt}
		\centering
		\subfloat[Data-Heavy]{
			\includegraphics[width=.508\linewidth]{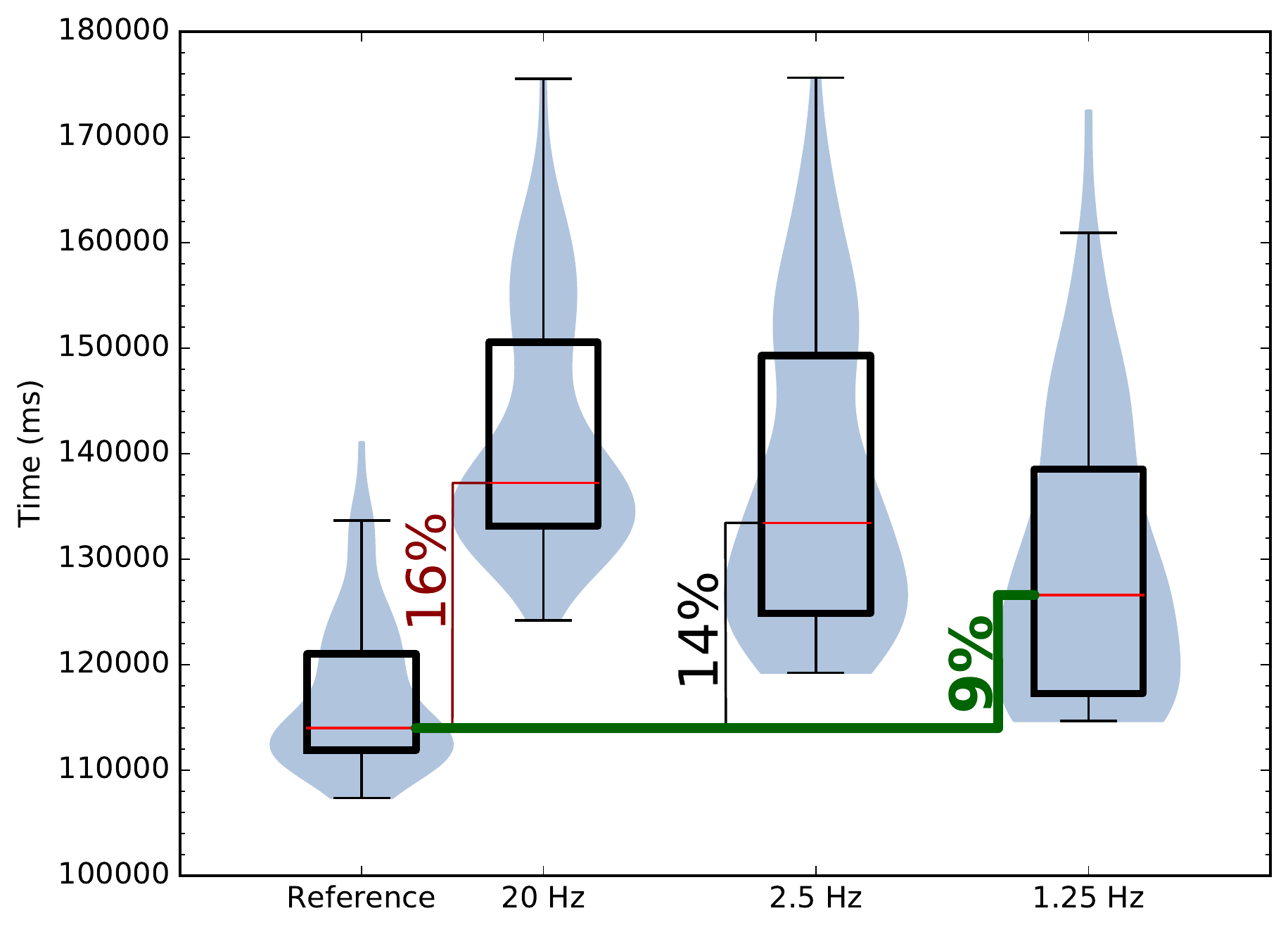}
		}%
		\subfloat[Very Data-Heavy]{
			\includegraphics[width=.492\linewidth]{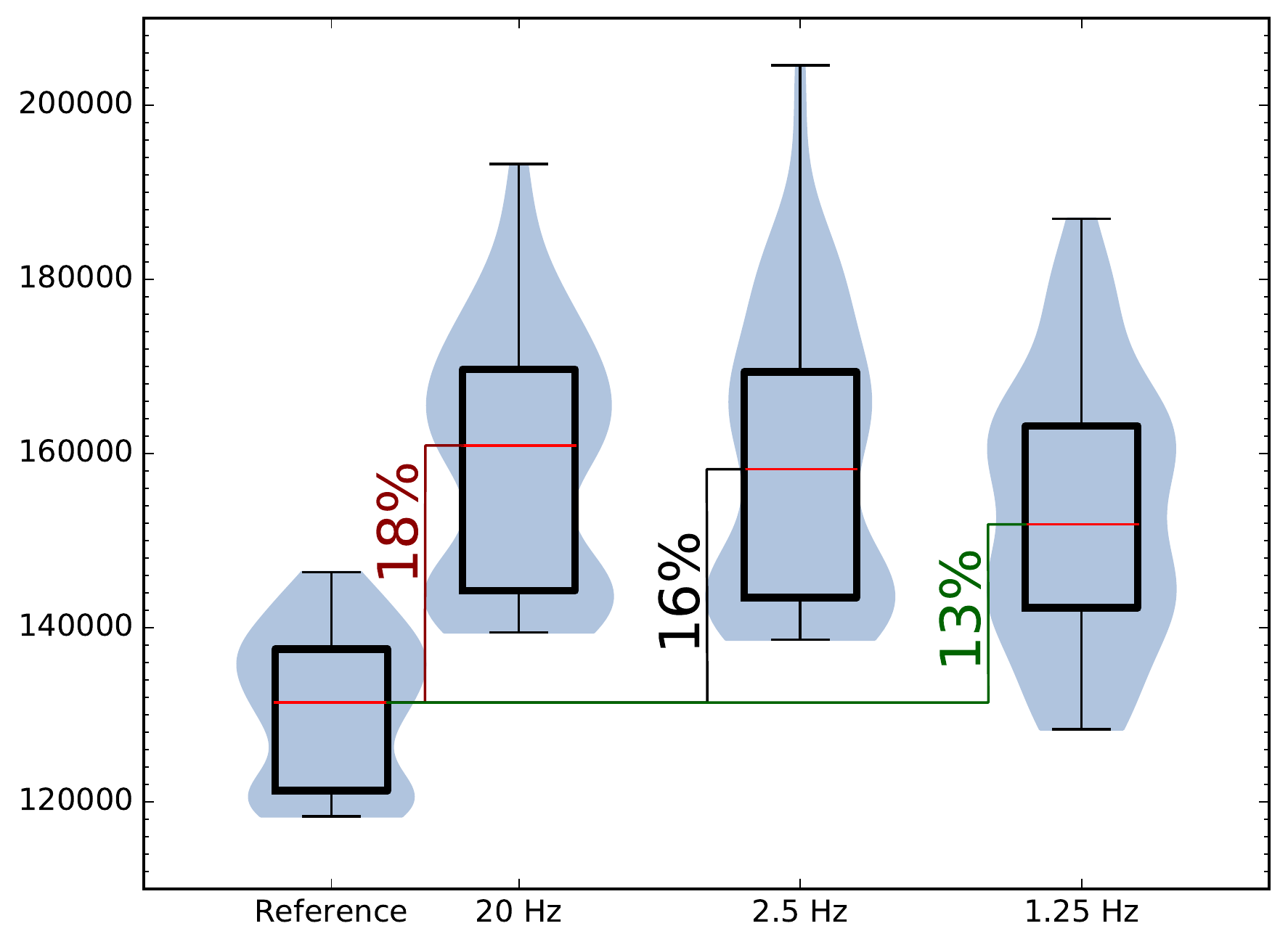}
		}%
		\\
		\subfloat[Balanced Compute-Heavy]{
			\includegraphics[width=.508\linewidth]{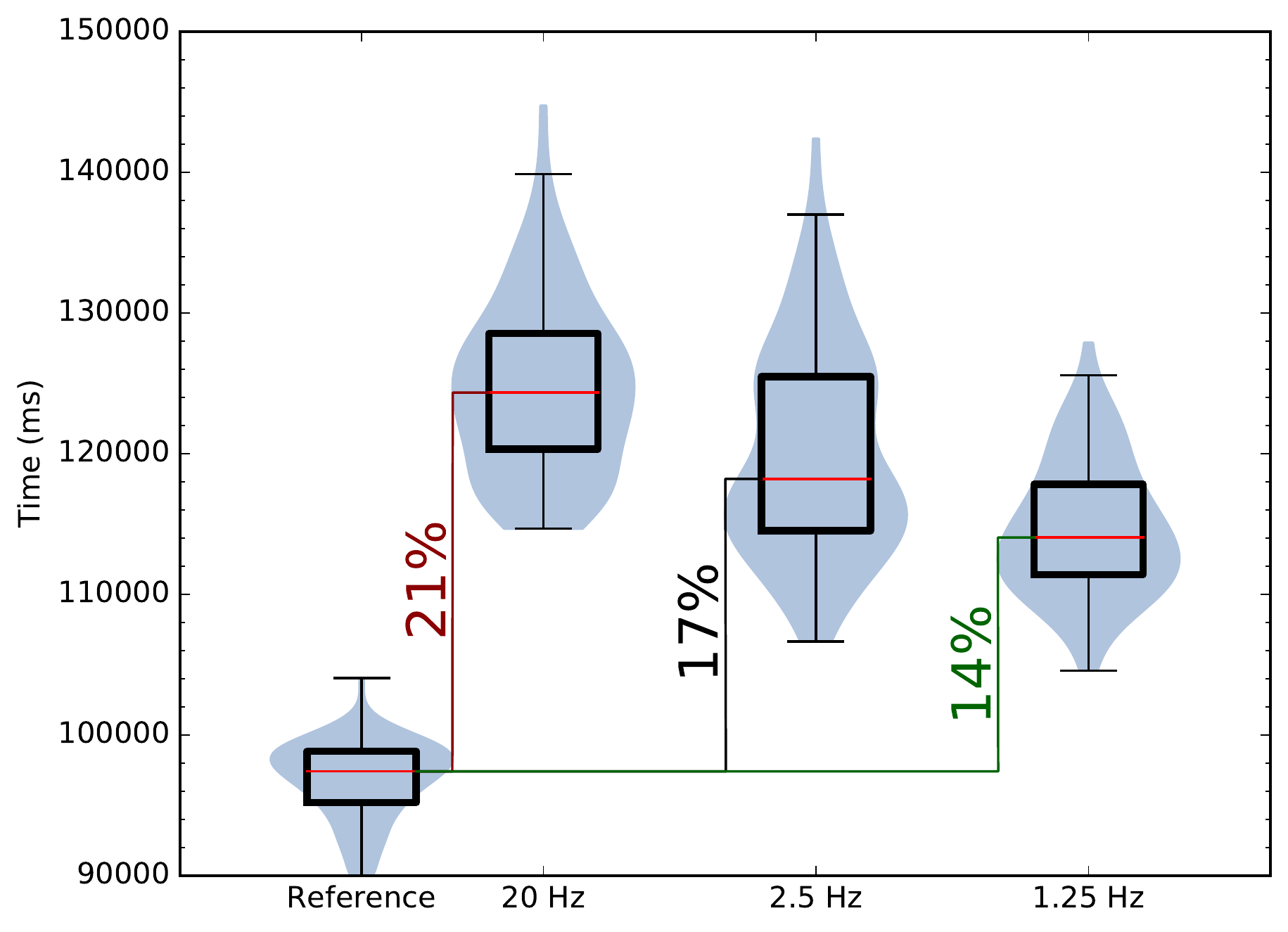}
		}
		\subfloat[Balanced Data-Heavy]{
			\includegraphics[width=.492\linewidth]{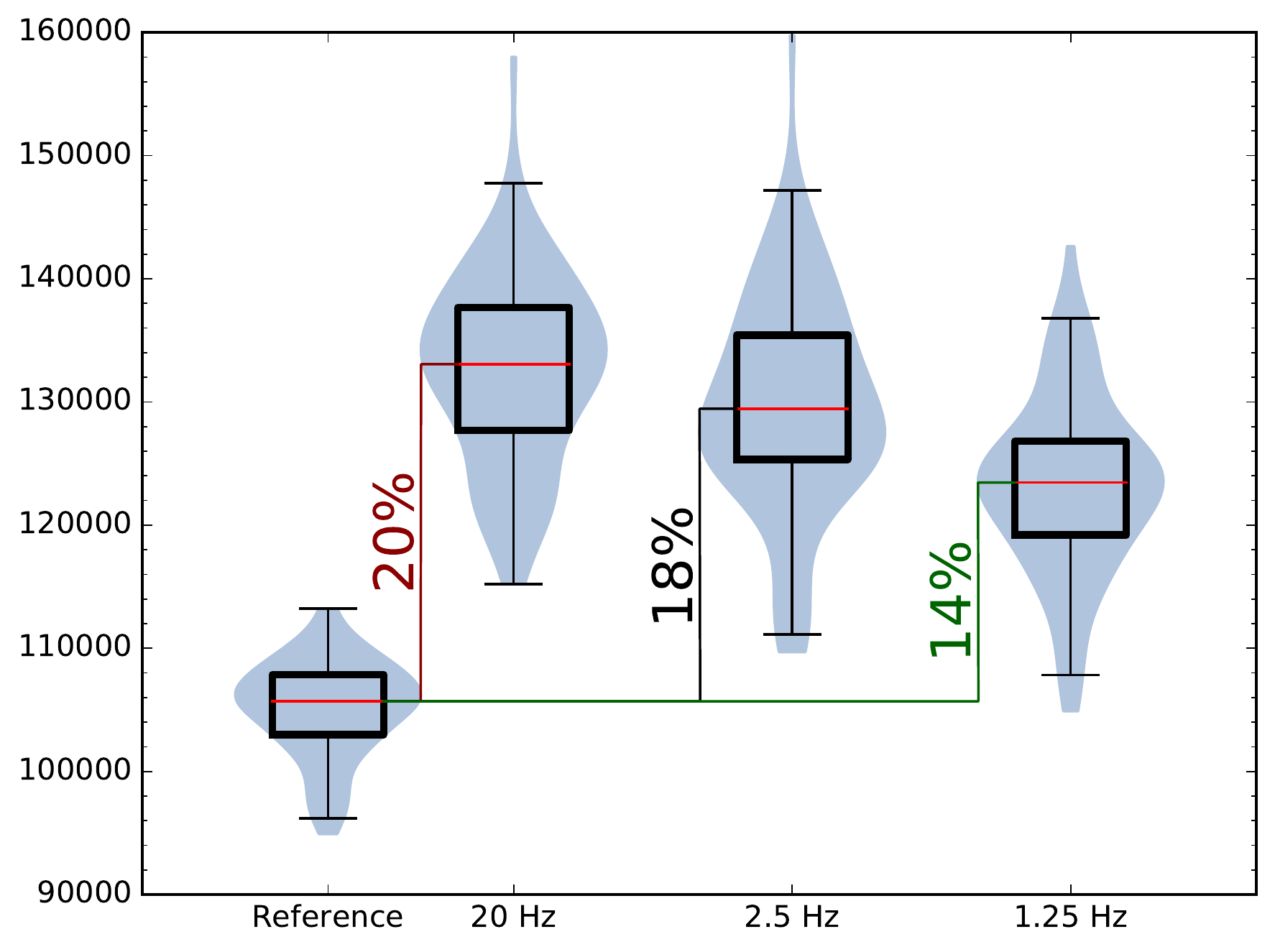}
		}
		\\
		\subfloat[Very Compute-Heavy]{
			\includegraphics[width=.508\linewidth]{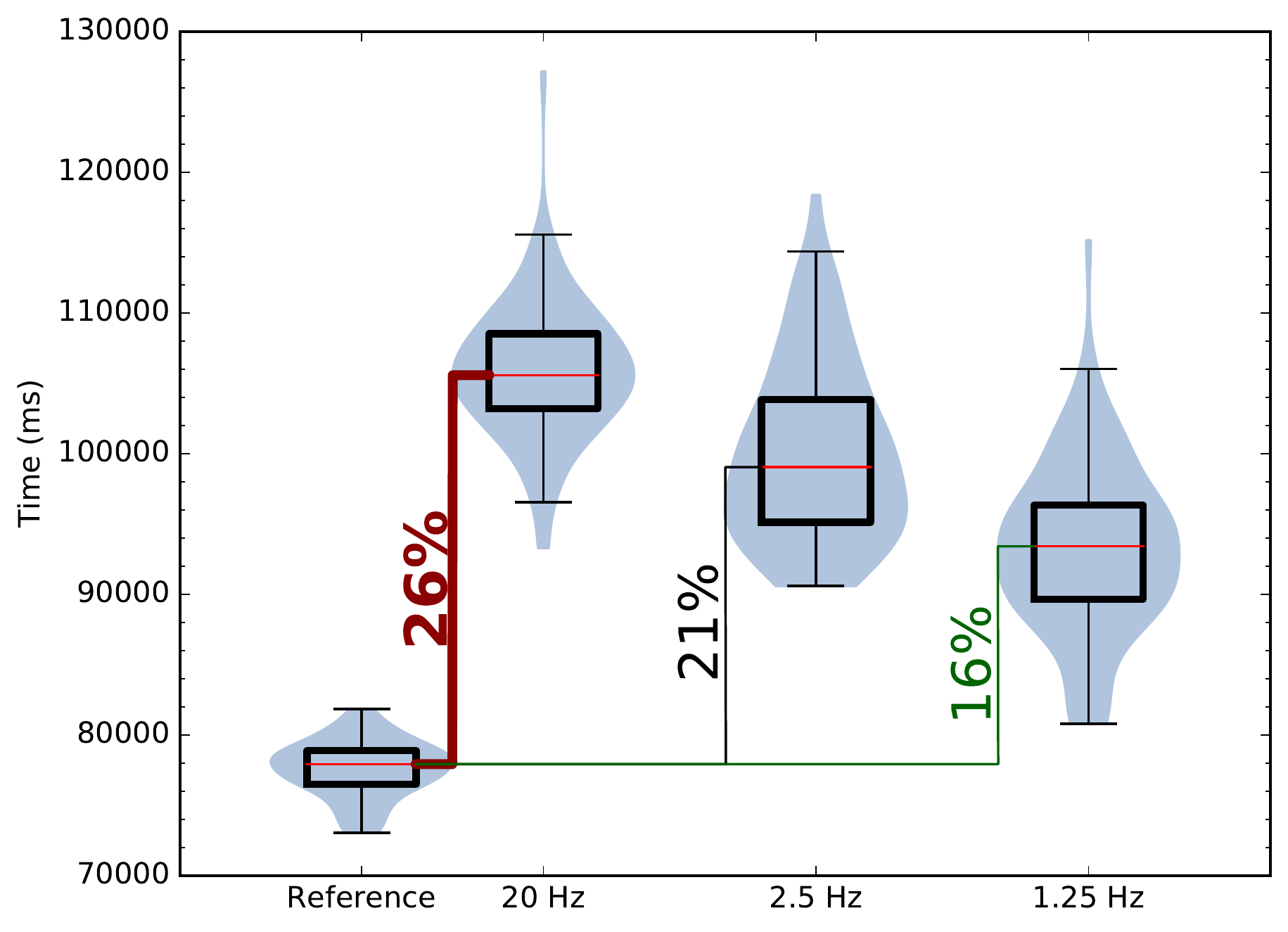}
		}
		\subfloat[Compute-Heavy]{
			\includegraphics[width=.492\linewidth]{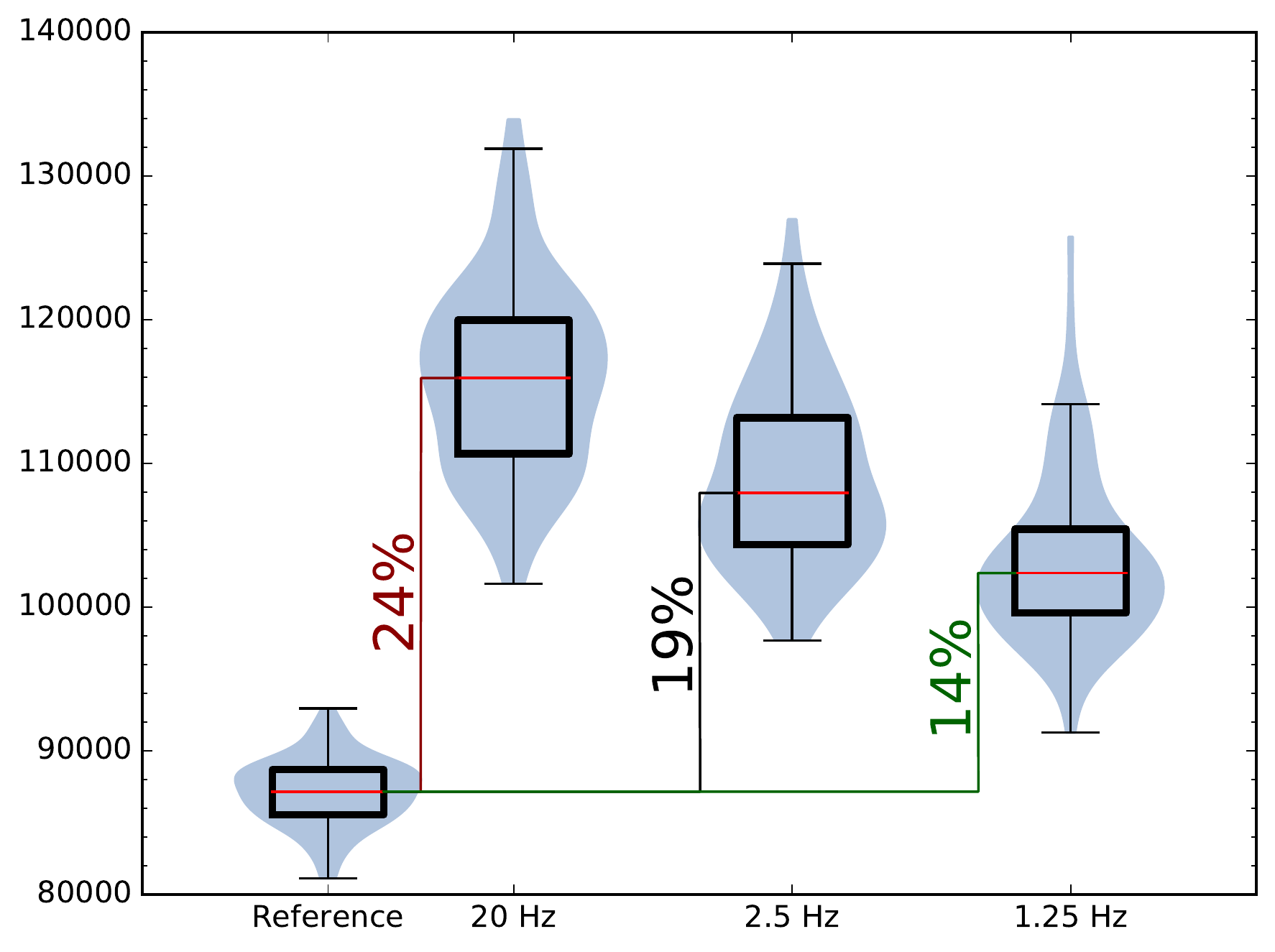}
		}
		\caption{Performance measurements for processing 600 frames with different checking frequencies and workloads.}
		\label{fig:performance}
	\end{figure}

	This naive implementation of our approach at the application level on Linux shows median-best performance degradation of 9\% and median-worst degradation of 26\%, which are also indicated in Figure \ref{fig:performance}a and e in bold.
	Across all test runs, we measured on average 80\% worst-case and 95\% best-case performance compared to the unprotected reference runtime.
	The violin plots -- shadows around the box-plots -- indicate the distribution of the measurements to depict the accumulation of the individual measurements.
	
	As expected, the performance varies depending on workload, with data-heavy tasks a-c showing better performance.
	This too was expected as the first stage's code consists mainly of function calls, integer operations, binary comparisons, and jumps.
	Drastically better performance can be expected in a more optimized implementation at the kernel level.
	To put these measurements into context, even a 50\% performance degradation on modern MPSoCs will offer a factor-of-5 performance increase over state-of-the-art radiation-hardened processor designs.
	Assuming an average performance degradation between 10\% and 20\%, our approach can thus allow a modern MPSoC to perform drastically better than comparable state-of-the-art solutions, while requiring no proprietary processor design, offering full software-control at a fraction of the development effort and costs.

	\section{RTOS based Implementation}
	
	After successfully implementing the previous stage and profiling the logic, we took this implementation and ported it to run bare-metal on a Cortex-A9 platform.
	Initially, we ported the concept to an RTOS to avoid the extreme performance overhead induced due to userland controlled threading.
	Thus, we implemented this concept in RTEMS, as it is one of the standard RTOS and the primary RTOS used within ESA associated space projects.
	This implementation was largely code-identical to the previous described implementation and also based upon POSIX threading.
	However, all costly low-level operations are now executed directly, no longer requiring system calls and context switches to kernel-mode.
	
	\subsection{Implementation Outline}
	Identical to prior section, except for bootstrapping being performed within the RTEMs executive.
	
	\subsection{Test Application}
	Identical to prior section.
	
	\subsection{Methodology and Test Setup}
	\begin{itemize}
		\item Cortex-A9 MPCore @667MHz
		\item Xilinx Zynq-XC7Z020 on Avnet Zedboard
		\item 256 MB RAM
	\end{itemize}
	
	\subsection{Results}
	Due to the extreme runtime to achieve significant and reproducible results, the quantity of measurements taken so far is not statistically significant.
	While a reduction of the test runtime would easily be possible, our objective at this stage is to retain the existing, well known test setup with its parameters.
	The results will be made available in a later version of this article.
%
%
	
	\section{Validation Memory Data Structures}
	This section contains additional information on the content and organization of \emph{validation memory}.
	The OS's view on the tile-local shared memory area (validation memory) is depicted in below listings.
	For the owner tile, this memory is writeable, other tiles have RO access.
	Each tile's local validation memory is mapped to the same memory address, whereas other tiles' validation memory is accessible at a fixed memory address via the shared interconnect.
	We also briefly illustrate the utilized voting-result bus described in the paper.
	
	\subsection{Tile-Global Public Data}
	\begin{lstlisting}[style=myc]
struct validation_memory {
/**
* flags for status and low-level signaling
*/
uint32_t status;
#define TILE_STATUS_FLAG_ACTIVE	(1)	// tile active?
#define TILE_STATUS_FLAG_ADDED	(1<<1)	// freshly added?
#define TILE_STATUS_FLAG_RESET	(1<<2)	// ongoing reset?
#define TILE_STATUS_FLAG_BROKEN	(1<<3)	// considered defunct?

/**
* flag register: this tile runs the following thread groups
*/
uint32_t member_of_thread_group;

/**
* flag register: disagreement with tileX
* See also "Result Bus"
*/
uint32_t disagree;

/**
* all the run-time information for our tile's threads
*/
struct thread_info tiles_threads [NUM_THREADS];

/**
* Global checkpoing frequency calculated as GCD of all active
* thread groups and tile groups. Each time a new thread group
* is added to this tile, we update this value to:
* 
* GCD(old_global_checkpoint_freq, thread->app->checking_freq)
*/
uint32_t global_checkpoint_freq;

/**
* all the thread meta information for our tile (dynamic part)
* dynamic information, is invalid if TILE_STATUS_FLAG_ADDED 
* or _BROKEN set!
*/
uint8_t dynamic [];
};
	\end{lstlisting}
	
	\subsection{Thread-level Data Structure (Validation Memory)}
	\begin{lstlisting}[style=myc]
struct thread_info {		
/**
* Checking intervall based upon the global checkpoint frequency
* Not during all checkpoints, a thread will be checked unless:
*
* thread_checkpoint_freq == global_checkpoint_freq
*
* otherwise:
*
* checkpoint_interval = thread_freq / global_checkpoint_freq
*/
int checkpoint_interval;

/**
* A check is performed for this thread, if this counter reaches
* 0, thus a check is pending. 
* 
* Gets set to checkpoint_interval each time this thread was
* checked during a checkpoint, and decremented each checkpoint
*/
int next_check;

/**
* flags for status and low-level signaling
*/
uint32_t status;
#define THREAD_STATUS_FLAG_CS_VALID	(1)	// Checksum rdy?
#define THREAD_STATUS_FLAG_FAILURE	(1<<1)	// crash/...

/**
* Checksum for comparison with and by siblings
* Current implementation uses CRC32 from zlib
*/
crc32_t csum;

/**
* length of the checksum data in bytes at *csum
*/
uint16_t len;

/**
* Size of the state buffer used by the sync() and update()
* checking routines.
*/
char data[STATE_BUFFER_SIZE_BYTES];
};
	\end{lstlisting}
	
	\section{A Demonstration MPSoC Design for a Development Board}
\begin{figure}[htb]
		\centering
		\includegraphics[width=0.9\linewidth]{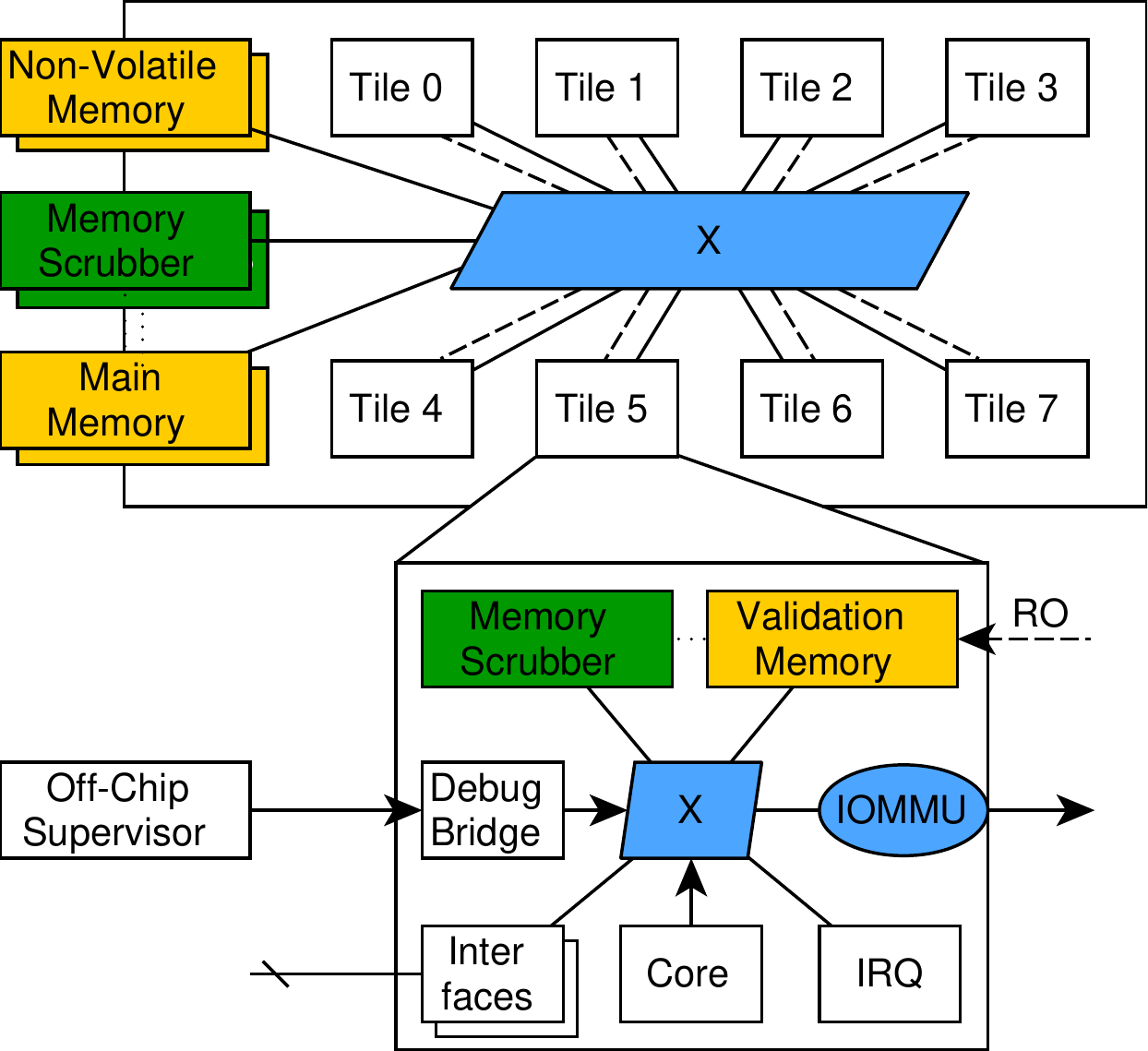}
		\caption{A simplified representation of the presented MPSoC with memory controllers highlighted in yellow, memory scrubbers in green, and the interconnect in blue. 
			An interconnect-bridge on each tile enables supervisor access.}
		\label{fig:mpsocidea}
	\end{figure}

	We developed a reproducible version of the the idealized MPSoC architecture presented in related works and depicted in Figure \ref{fig:mpsocidea} for reference.
	It utilizes Xilinx Microblaze processor cores instead or ARM cores as these and relevant peripheral IP-cores are more easily available to the community as part of Xilinx' Vivado Design Suite.
	While the idealized version omitted several implementation related details and in turn includes IP-cores not openly available, this reduced MPSoC design can be and was implemented using the Xilinx library-IP.

	The subsequently depicted design supports widely available standard FPGA development boards, which of course do not possess redundant, independent DDR-memories.
	Logic placement and fabric utilization are based upon the Xilinx Virtex VCU118 Development Kit, on which this design was implemented successfully.
	It can support the full functionality of the approach discussed in related work without requiring proprietary IP which will not be available to other researchers, and instead e.g. Xilinx Microblaze cores can be used.
	
\begin{figure*}[htb]
		\centering
		\includegraphics[width=0.7\linewidth]{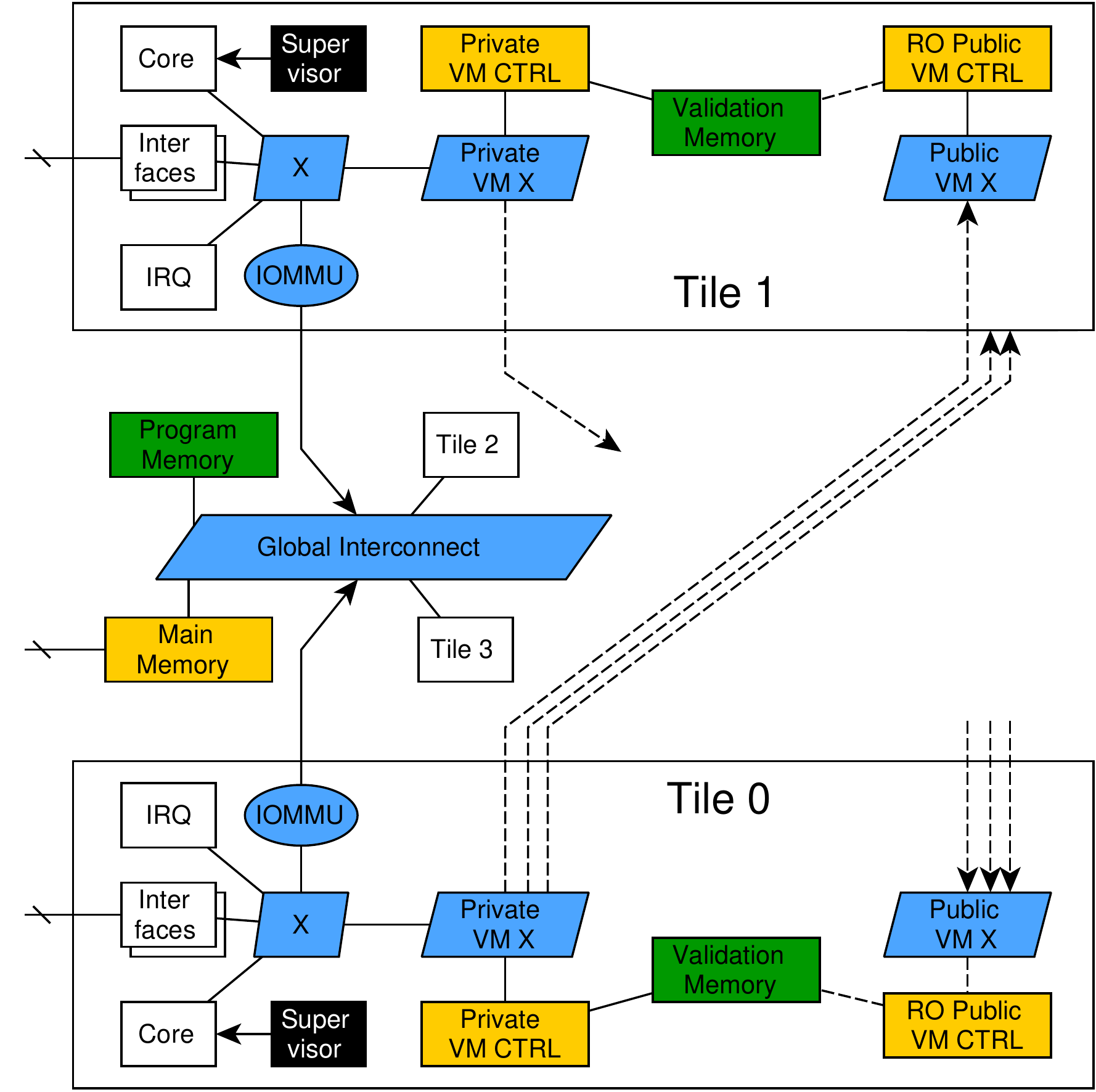}
		\caption{A more realistic version of the MPSoC depicted in Figure \ref{fig:mpsocidea}, including dedicated interconnects for validation memory and split read-only (tile-public) and tile-private memory controllers.}
		\label{fig:tile}
	\end{figure*}

\begin{figure*}[p]
		\centering
		\includegraphics[angle=90,width=0.7\linewidth]{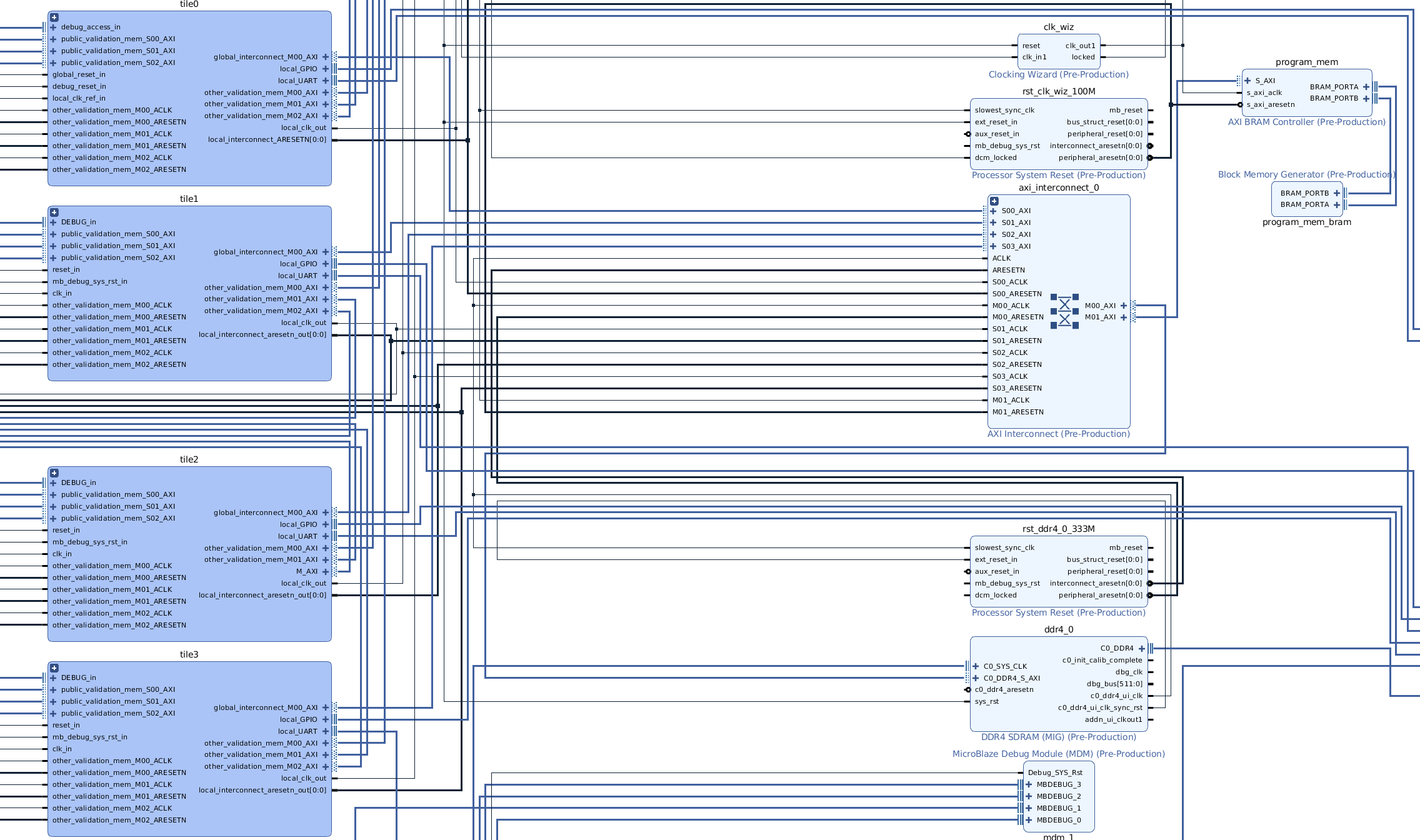}
		\caption{Global interconnect \& shared resources of the previously depicted MPSoC.}
		\label{fig:interconnect}
	\end{figure*}

\begin{figure*}[p]
		\centering
		\caption{A close-up view of a tile as depicted in Figure \ref{fig:interconnect}. The additional interconnects for validation memory are necessary to allow Vivado-side management of the address space without requiring IOMMUs for each memory.
		Furthermore, this setup reduces complexity of the shared global interconnect, and improves timing by simplifying logic placement.}
		\includegraphics[angle=90,width=0.65\linewidth]{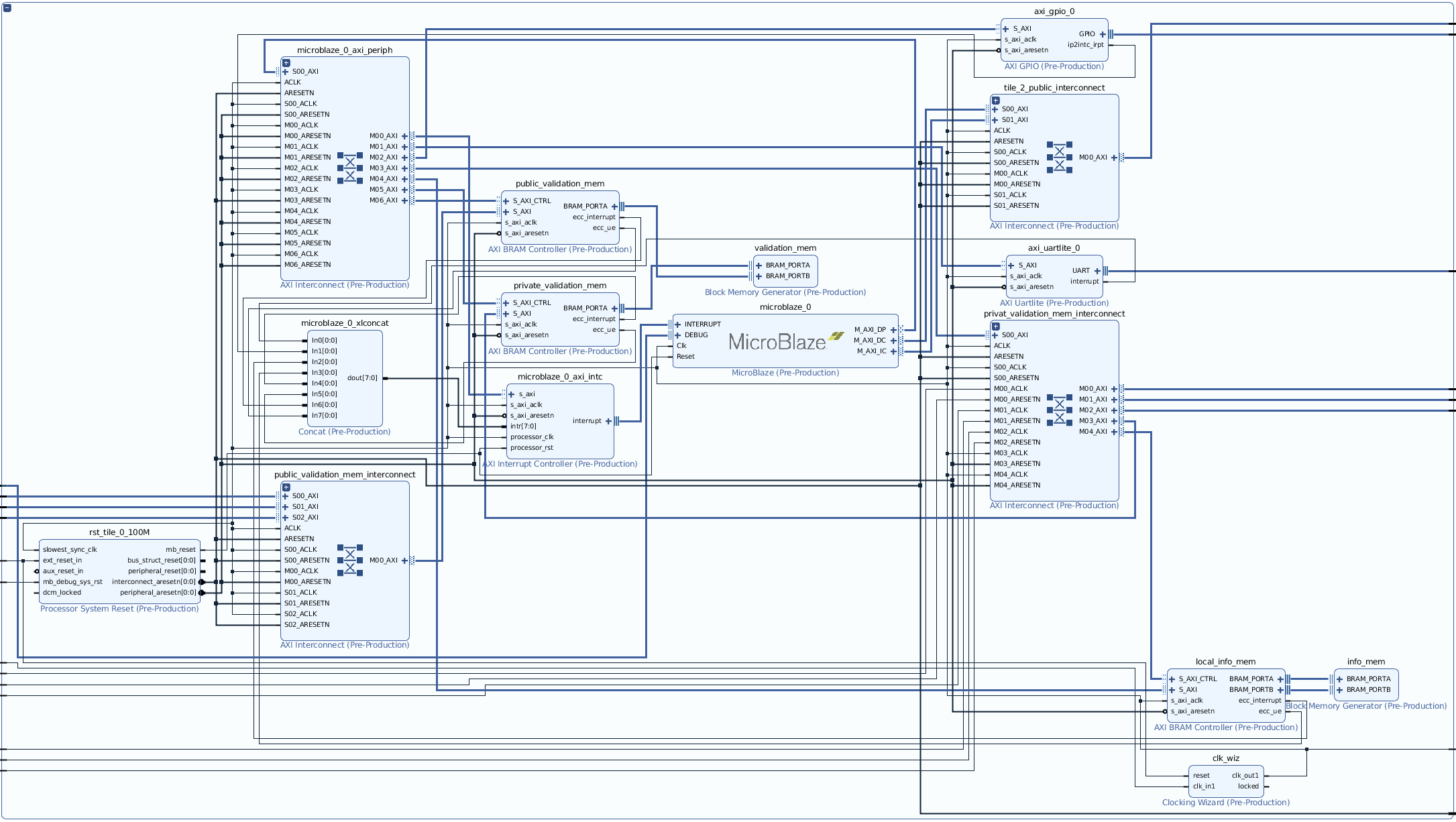}
		\label{fig:vivadotile}
	\end{figure*}

	Figure \ref{fig:tile} contains a simplified overview of the architecture implemented in the figures at the end of this supplementary article.
	A high-level depiction of the demo-MPSoC design's global interconnect, shared memory taken from the Vivado Design Studio Suite is provided in Figure \ref{fig:interconnect}.
	Please note that in contrast to the full MPSoC implementation, this demo architecture utilizes a single DDR4 memory controllers to drive the on-device DDR memory, and uses on-chip BRAM as program memory.
	This is unsuitable for a truly fault-tolerant setup, and program memory and the DDR4 controller are drastically larger than the memory controllers commonly used for space applications.
	On-chip BRAM can directly be programmed and manipulated by the Vivado toolchain using MicroBlaze debug functionality, and used here to simplify the demo.
	A close-up of a tile is depicted in Figure \ref{fig:vivadotile}, and includes an interrupt controller, a UART interface, validation memory and a local scratchpad memory-block implemented in BRAM and a GPIO controller to signal agreement between tiles.
	
\begin{figure*}[p]
		\centering
		\caption{Logic placement of the MPSoC depicted in Figures \ref{fig:interconnect} and \ref{fig:vivadotile} for a VCU118 development board running at 300MHz clock speed with dedicated clock domains for each tile and shared components.
		Tile 0-3 are depicted in green, red, yellow and pink, respectively.
		The global interconnect as well as other shared global components are shown in white, while the DDR4 memory controller is depicted in deep blue.
		Turquoise elements and lines indicate on-chip BRAM used as program memory for the demo-MPSoC.}
		\includegraphics[width=0.60\linewidth]{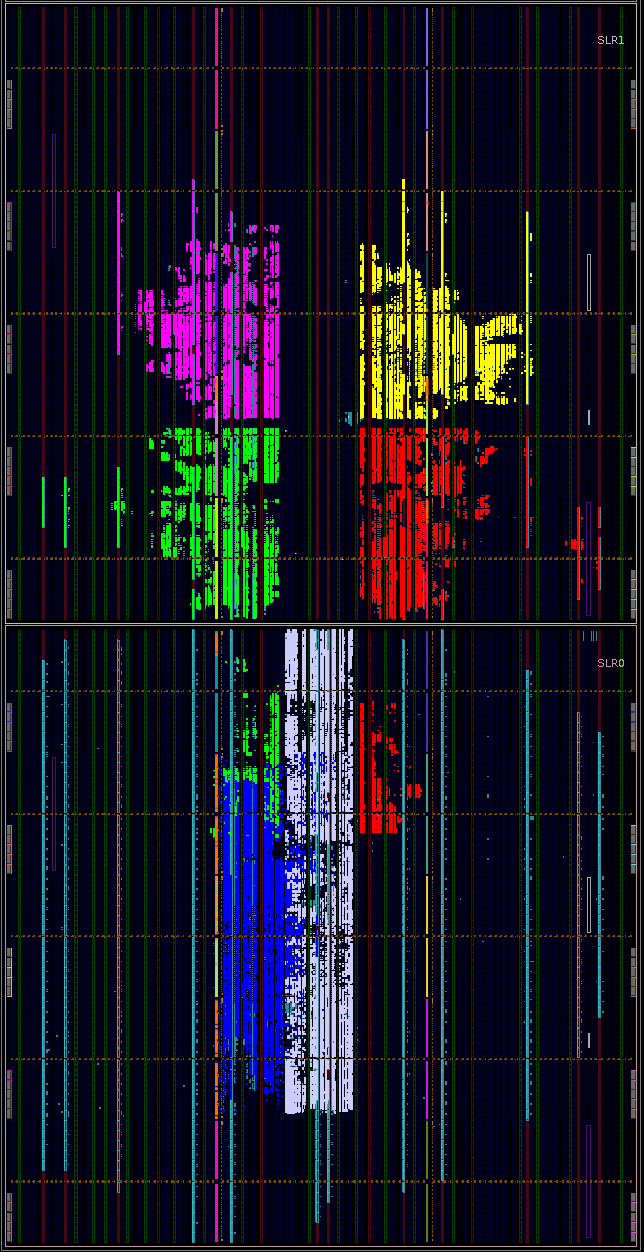}
		\label{fig:usagevisual}
	\end{figure*}

\begin{figure*}[p]
		\centering
		\caption{Resource Utilization for the MPSoC implementation on a VCU118 development board.}
		\includegraphics[angle=90,width=0.87\linewidth]{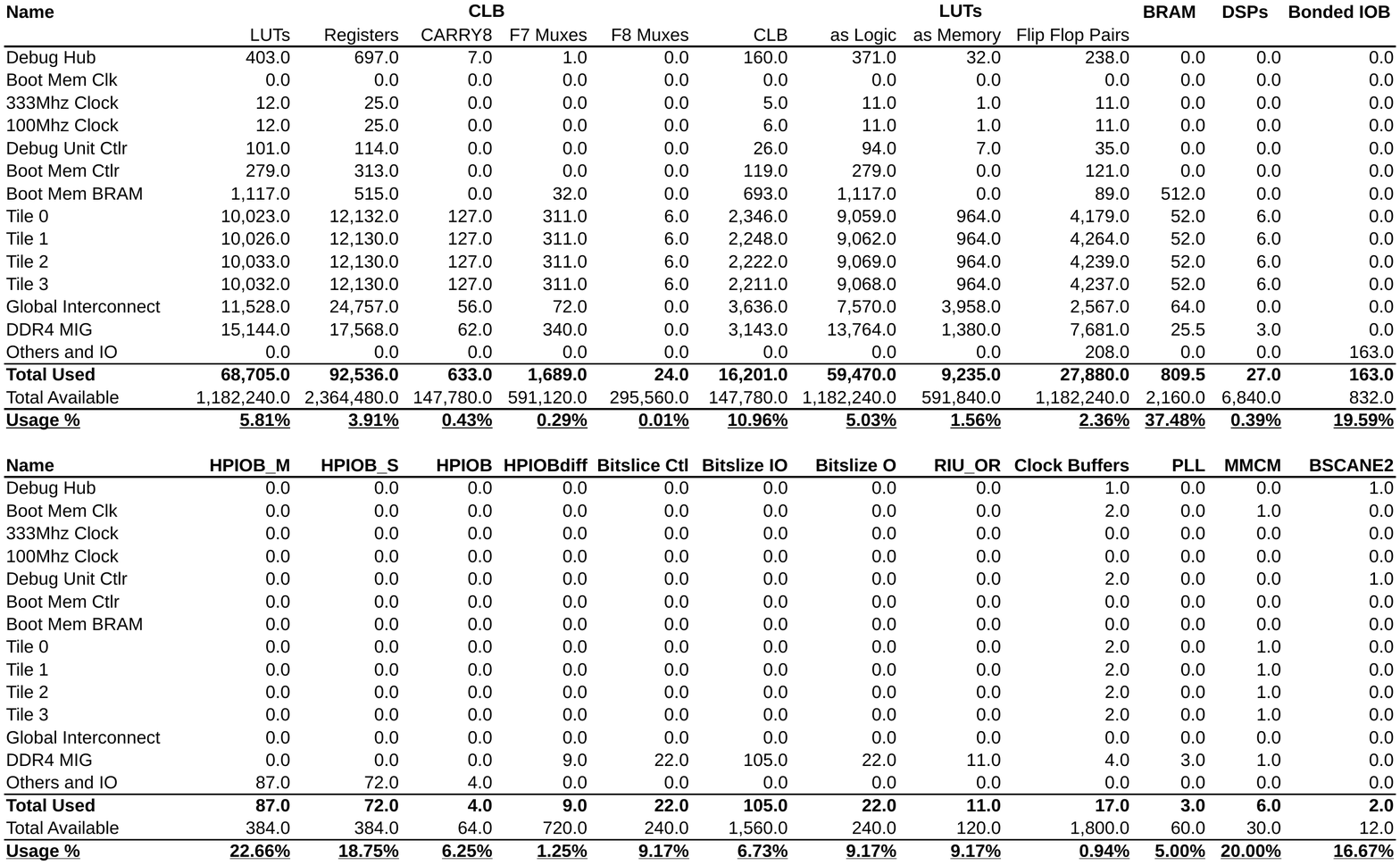}
		\label{fig:usage}
	\end{figure*}

	Logic placement with a VCU118 development kit is depicted in Figure \ref{fig:usagevisual}.
	Notice the placement of the DDR4-IO pads highlighted in orange which are development-kit imposed constraints.
	The figure only depicts the bottom half of the FPGA fabric, with the top half being unused and cut-off to improve readability.
	Different MPSoC components are colorized with color-indications being provided in the figure caption.
	The resource utilization report for this design is given in tabular form in Figure \ref{fig:usage}.
	
	External supervision is implemented through a set of external GPIO pins, and we can utilize a combination of Xilinx ChipScope and the Microblaze debug unit (MDB) to receive consistency decisions.
	To illustrate the general idea of how the output bus looks like, we provide a VHDL code listing.
	In our Vivado-based implementation, we of course utilize the library MUX.
	Alternatively, this bus can be implemented using the GPIO pins of a COTS MPSoC or simply as depicted below in memory.
	\begin{lstlisting}[style=myvhdl]
	entity result_bus_mux is
	Port (
	SEL	: in	STD_LOGIC;
	A	: in	STD_LOGIC_VECTOR (Tile_Count downto 0);
	B	: in	STD_LOGIC_VECTOR (Tile_Count downto 0);
	C	: in	STD_LOGIC_VECTOR (Tile_Count downto 0);
	--	...	...	...
	O	: out	STD_LOGIC_VECTOR (Tile_Count downto 0)
	);
	end result_bus_mux;
	\end{lstlisting}
	This bus can be utilized by each tile and in the C-based implementation accessed as a variable as depicted in below listing.
	
	\begin{lstlisting}[style=myc]
	/**
	* flag register: disagreement with tileX
	*/
	uint32_t disagree;
	\end{lstlisting}
	
	Supervision for the presented demo-MPSoC can also be implemented on-chip to implement a fully functional setup to simulate the concept.
	This can be achieved by adding a minimal Microblaze system to the MPSoC which has access to either each tile's GPIO interface and validation memory or can utilize the additional IP-core \emph{information memory} depicted in from Figure \ref{fig:vivadotile}.
	
	\section{Next Steps}
	
	As a next step, we plan to replace the global crossbar-based AXI interconnect with a NoC-based system.
	This could further be used to implement interconnect-level fault-tolerance by assuring message consistency and fail-over through replication.
	
	\bibliographystyle{IEEEtran}
	\bibliography{performance}
\end{document}